\documentclass[12pt]{article}
\usepackage{amsmath}
\usepackage{graphicx}
\usepackage{amsfonts}
\usepackage{amssymb}

\begin{document}

\begin{center}
{\huge Nucleon-nucleon interaction:}

{\huge Central potential and pion production}
\end{center}

\ 

\begin{center}
C. M. MAEKAWA$^{\dagger}$,\ \ J. C. PUPIN$^{\ast\,\ddagger}$\ \ and\ \ M. R.
ROBILOTTA$^{\ddagger}$
\end{center}

\ 

\begin{center}
$^{\dagger}${\small Kellogg Radiation Laboratory, 106-38}\linebreak
{\small California Institute of Technology, Pasadena, California
91125}\linebreak e-mail: \textit{maekawa@krl.caltech.edu}

$^{\ddagger}${\small Nuclear Theory and Elementary Particle Phenomenology
Group}\linebreak {\small Instituto de F\'{\i}sica, Universidade de S\~{a}o
Paulo}\linebreak {\small C.P. 66 318, 05315-970, S\~{a}o Paulo, SP,
Brazil}\linebreak e-mails: \textit{pupin@if.usp.br}\ and\ \textit{robilotta@if.usp.br}

$^{\ast}${\small Instituto de F\'{\i}sica Te\'{o}rica, Universidade Estadual
Paulista}\linebreak {\small 01405-900, Rua Pamplona 145, S\~{a}o Paulo, SP, Brazil}
\end{center}

\ 

\

\baselineskip=0.7cm

\begin{abstract}

We show that the tail of the chiral two-pion exchange nucleon-nucleon
potential is proportional to the $\pi N$ scalar form factor and discuss how it
can be translated into effective scalar meson interactions. We then construct
a kernel for the process $NN\rightarrow\pi NN$, due to the exchange of two
pions, which may be used in either three body forces or pion production in
$NN$ scattering. Our final expression involves a partial cancellation among
three terms, due to chiral symmetry, but the net result is still important. We
also find that, at large internucleon distances, the kernel has the same
spatial dependence as the central $NN$ potential and we produce expressions
relating these processes directly.

\ 

\ 

\ 

\ 

\ 

\ 


\end{abstract}

\newpage


\section{Introduction}

The description of pion production in nucleon-nucleon ($NN$) interactions near
threshold is a traditional problem in hadron physics. In recent times the
interest in it was renewed, due to a wealth of precise experimental data:
$np\rightarrow d\pi^{0}$ \cite{Hut+91}, $pp\rightarrow pp\pi^{0}$
\cite{Mey+92,Bond+95}, $pp\rightarrow d\pi^{+}$ \cite{Droch+96,Heimb+96},
$pp\rightarrow pn\pi^{+}$ \cite{Daeh+95}. On the theoretical side, the
availability of chiral perturbation theory (ChPT) allowed the problem to be
tackled in a systematic manner. However, in spite of all the effort made, a
satisfactory picture is still not available.

There are two classes of interactions involved in this process, associated
with either nucleon correlations or the emission of the external pion. In the
procedure developed by Koltun and Reitan \cite{Kol+66}, these interactions are
encompassed in wave functions and interaction kernels. The former correspond
to solutions of the Schr\"{o}dinger equation with realistic potentials,
whereas the latter are described by models based on Feynman diagrams.

\begin{figure}
[h]
\begin{center}
\includegraphics[
height=1.5515in,
width=5.9629in
]%
{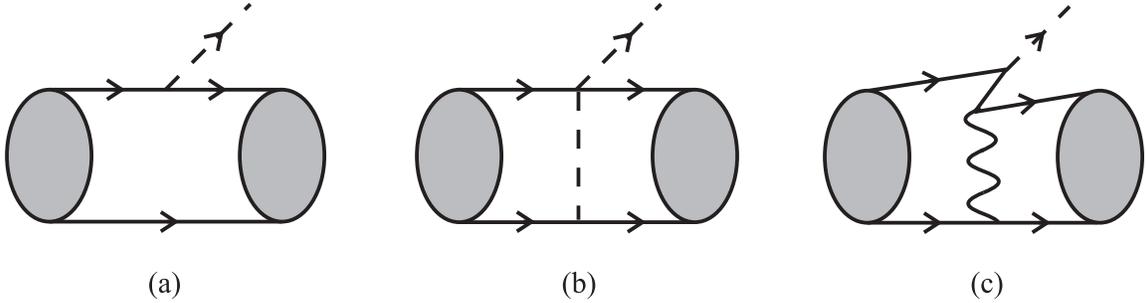}%
\caption{Contributions to the process $NN\rightarrow\pi NN$: (a) impulse, (b)
rescattering and (c) z-graphs; nucleons, pions and heavier mesons are
represented by solid, dashed and wavy lines.}%
\label{Fig.1}%
\end{center}
\end{figure}

In the discussion of the production kernel, one usually distinguishes between
long and short range contributions. The former are shown in Fig.\ref{Fig.1},
where the first diagram represents the impulse approximation and the second,
the pion rescattering term. These two processes were considered by Koltun and
Reitan \cite{Kol+66} in their description of the $\pi^{0}$ channel, but the
corresponding cross section proved to underestimate recent data
\cite{Hut+91,Mey+92,Bond+95} by a factor 5 \cite{Mill+91}. The rescattering
term used in that work came from on shell $\pi N$ amplitudes, whereas the pion
exchanged in diagram 1b is off-shell. Models which take pion virtuality into
account enhances the cross section and tend to reduce underprediction
\cite{Hach+78,Efro+85,Hern+95,Han+95}. Heavy baryon ChPT calculations
\cite{Kolck+96,Mill+96,Caroch+99} also stressed the importance of this
rescattering term at leading order. However, in these works the rescattering
and impulse terms came out with opposite signs and the net result was again
smaller than in phenomenological calculations \cite{Spe+98}. Hence other
mechanisms are needed to improve the description.

The next natural step concerns shorter range interactions, especially those
involving two pions. The treatment of uncorrelated two-pion exchange is rather
complex and, in the case of pion production, this part of the interaction has
been described by effective heavy meson exchanges. Their contributions
correspond to the last diagram of Fig.\ref{Fig.1}, known as z-graph, since
positive frequency nucleon propagation, already included in the wave function,
is subtracted. The inclusion of $\sigma$, $\omega$ and $\rho$ mesons, either
explicitly \cite{Han+95} or into a general axial current density
\cite{Riska+93} gave rise to good fits to the $pp\rightarrow pp\pi^{0}$ cross
section at threshold. However, the study of relativistic effects in $\pi^{0}$
production \cite{Adam+97} demonstrated that z-graph effects are small and
hence heavy mesons in ChPT \cite{Kolck+96,Mill+96} do not give rise to the
required increase in cross sections. Extension to other channels ($\pi^{+}$
and $\pi^{-}$), with inclusion of nucleon resonances
\cite{Caroch+99,Han+98,Riska+99} also did not improve the situation.%

Some time ago, Coon, Pe\~{n}a and Riska \cite{CPR} produced a three-body
potential based on the exchanges of a pion and a scalar meson, which proved to
be able to reduce the gap between theory and experiment for the binding energy
of trinuclei. Later on, we derived an equivalent result, using a non-linear
Lagrangian, which included an effective chiral scalar meson coupled to
nucleons \cite{MRtbf}. In that work, the effective field was designed to
simulate the two pion exchange potential. We stress that the exchange of two
uncorrelated pions, formulated in the framework of chiral symmetry and
including delta degrees of freedom, explains quite well the tail of the
scalar-isoscalar nucleon-nucleon potential \cite{ORK,RR,K+} and there is no
need at all for a true scalar meson to describe that channel. On the other
hand, the treatment of uncorrelated two pion exchange requires the calculation
of many Feynman diagrams and, in problems where one is more concerned with
simplicity than with fine details, it may be useful to replace all processes
associated with the scalar-isoscalar channel by a single effective field. In
this conceptual framework, our Lagrangian gave rise to a strong
pion-scalar-nucleon contact interaction, that corresponds to the kernel for
the reaction $NN\rightarrow\pi NN$ due to the exchange of two pions. This
kernel was then applied to the $\pi^{0}$ and $\pi^{+}$ production channels and
theoretical results were found to be comparable to the experimental ones
\cite{MdR}. Recently, calculations based on both relativistic \cite{BKM} and
heavy baryon ChPT \cite{Sato} dealt with such a transition operator and large
contributions were again found, involving several cancellations.

The purpose of the present work is two fold. In Sec.II, we discuss the
relationship between the actual tail of the two-pion exchange potential and
that provided by an effective scalar meson. The pion-production kernel is
considered in Sec.III and, in order to determine the role of chiral
cancellations, we study the leading term, constructed by means of the $\pi
N\rightarrow\pi N$ and $\pi N\rightarrow\pi\pi N$ subamplitudes. Our results
are indeed based on a partial cancellation, involving three large factors. We
also find that, at large internucleon distances, the kernel has the same
spatial dependence as the central $NN$ potential and hence, in Sec.IV, we
produce expressions relating these interactions directly. Finally, in Sec.V we
present a summary and conclusions.


\section{Central Potential}

\begin{figure}
[h]
\begin{center}
\includegraphics[
height=1.6155in,
width=5.0531in
]%
{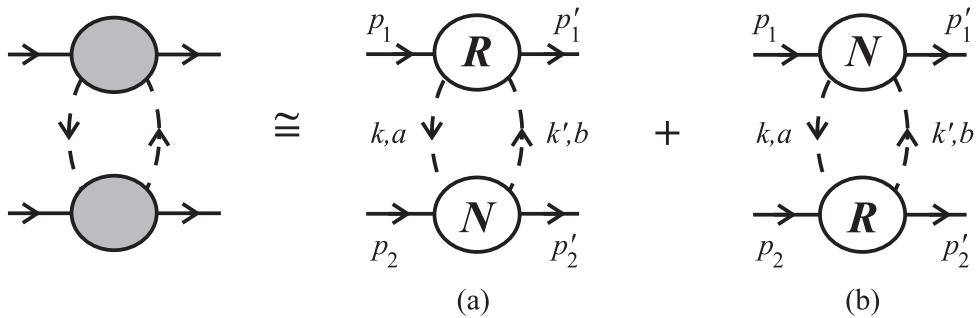}%
\caption{Leading contributions to the process $NN\rightarrow NN$.}%
\label{Fig.2}%
\end{center}
\end{figure}

The two-pion exchange potential (TPEP) is closely related to $\pi N$
scattering. The isoscalar amplitude for the process $NN\rightarrow NN$ is
represented in Fig.\ref{Fig.2} and given by%
\begin{equation}
\mathcal{T}^{S}\,=\,-\,\frac{i}{2!}\int\frac{d^{4}Q}{(2\pi)^{4}}%
\;\frac{3\;[T^{+}]^{(1)}\;[T^{+}]^{(2)}}{(k^{2}-\mu^{2})\,(k^{\prime\,2}%
-\mu^{2})}\;, \label{C1}%
\end{equation}
where $T^{+}$ is the isospin symmetric part of amplitude for the process
$\pi^{a}N\rightarrow\pi^{b}N$ and $Q=(k^{\prime}+k)/2$.%

In recent times, chiral symmetry has been systematically applied to this
problem and one has learned \cite{ORK, RR, K+} that it is convenient to
separate $T^{+}$ into a contribution $T_{N}$, due only to pion-nucleon
interactions, and a remainder $T_{R}$, involving other degrees of freedom. One
then writes symbolically $[T^{+}]=[T_{N}^{+}]+[T_{R}^{+}]$ for each nucleon
and the potential is then proportional to $[T^{+}]^{(1)}[T^{+}]^{(2)}%
=[T_{N}^{+}]^{(1)}[T_{N}^{+}]^{(2)}+\left\{  [T_{N}^{+}]^{(1)}[T_{R}%
^{+}]^{(2)}+[T_{R}^{+}]^{(1)}[T_{N}^{+}]^{(2)}\right\}  +[T_{R}^{+}%
]^{(1)}[T_{R}^{+}]^{(2)}$. The numerical study of these contributions has
shown that the term within curly brackets is largely dominant \cite{RR} and
due to the subamplitudes
\begin{align}
T_{N}^{+} &  =\frac{g^{2}}{m}\;\bar{u}\,\left\{  1-\left[  \frac{m}%
{(p+k)^{2}-m^{2}}-\;\frac{m}{(p-k^{\prime})^{2}-m^{2}}\right]  \,%
\!\!\not\!%
Q\right\}  \,u\;,\label{C2}\\[0.16in]
T_{R}^{+} &  =\bar{A}^{+}(\nu=0,t=4\mu^{2})\;\bar{u}\,u=\frac{\alpha_{00}^{+}%
}{\mu}\;\bar{u}\,u\;,\label{C3}%
\end{align}
where $g$ is the $\pi N$ coupling constant, $m$ is the nucleon mass and the
bar over the isospin even $\pi N$ subamplitude $A^{+}$ indicates the
subtraction of the pseudoscalar Born term. The constant $\alpha_{00}^{+}$ may
be expressed as combination \cite{RR} of $\pi N$ subthreshold coefficients
\cite{H83}. The leading contribution to $\mathcal{T}^{S}$ is written as
\begin{equation}
\mathcal{T}^{S}\;\cong\;3\,\frac{\alpha_{00}^{+}}{\mu}\;[\bar{u}\,u]^{(1)}%
\;\left[  \Gamma_{N}^{+}\right]  ^{(2)}\;+\;\;(1\leftrightarrow2)\;,\label{C4}%
\end{equation}
where
\begin{equation}
\left[  \Gamma_{N}^{+}\right]  ^{(2)}=-\;\frac{i}{2}\int\frac{d^{4}Q}%
{(2\pi)^{4}}\;\frac{\left[  T_{N}^{+}\right]  ^{(2)}}{[(Q\!-\!\Delta
/2)^{2}\!-\!\mu^{2}]\;[(Q\!+\!\Delta/2)^{2}\!-\!\mu^{2}]}\;,\label{C5}%
\end{equation}
with $\Delta=(k^{\prime}-k)$. Using Eq.(\ref{C2}), we have
\begin{align}
&  \left[  \Gamma_{N}^{+}\right]  ^{(2)}=-\,\frac{i}{2}\,\frac{g^{2}}%
{m}\,\left\{  \left[  \bar{u}\,u\right]  ^{(2)}\int\frac{d^{4}Q}{(2\pi)^{4}%
}\;\frac{1}{[(Q\!-\!\Delta/2)^{2}\!-\!\mu^{2}]\;[(Q\!+\!\Delta/2)^{2}%
\!-\!\mu^{2}]}\right.  \nonumber\\[2mm]
&  -\left.  \left[  \bar{u}\,\gamma_{\mu}\,u\right]  ^{(2)}\int\frac{d^{4}%
Q}{(2\pi)^{4}}\;\frac{2m\;Q^{\mu}}{[(Q\!-\!\Delta/2)^{2}\!-\!\mu
^{2}]\;[(Q\!+\!\Delta/2)^{2}\!-\!\mu^{2}]\;[Q^{2}\!+\!2mV_{2}\!\cdot
\!Q\!-\!\Delta^{2}/4]}\right\}  \nonumber\\[2mm]
&  =\frac{1}{2}\,\frac{g^{2}}{m}\,\frac{1}{(4\pi)^{2}}\,\left[  J_{c,c}%
(t)-J_{c,sN}^{(1)}(t)\right]  \;\left[  \bar{u}\,u\right]  ^{(2)}\,.\label{C6}%
\end{align}

In deriving this result we used $V_{2}=(p_{2}^{\prime}+p_{2})/2m$ and the
symmetry of the integrand under $Q\rightarrow-Q$. The functions $J$, defined
in Ref.\cite{MRR99}, are given by
\begin{align}
J_{c,c}(t)  &  =C(d,\Lambda)-\mu^{2}\int_{0}^{1}d\alpha\int_{0}^{1}%
\frac{d\beta}{\beta}\;\frac{\lambda^{2}}{t-\lambda^{2}\,\mu^{2}}%
\,,\label{C7}\\[0.4cm]
J_{c,sN}^{(1)}(t)  &  =-\,2m^{2}\int_{0}^{1}d\alpha\;\frac{1-\alpha}{\alpha
}\int_{0}^{1}d\beta\;\frac{1-\beta}{\beta}\;\frac{1}{t-\eta^{2}\,\mu^{2}}\;,
\label{C8}%
\end{align}
with $t=\Delta^{2}$ and
\begin{align}
\lambda^{2}  &  =1/\left[  \alpha(1-\alpha)\beta\right]  \;, \label{C9}%
\\[0.2cm]
\eta^{2}  &  =\left[  (1-\alpha)^{2}(1-\beta)^{2}\,m^{2}/\mu^{2}%
+1-(1-\alpha)(1-\beta)\right]  \,/\,\left[  \alpha(1-\alpha)\beta\right]  \,.
\label{C10}%
\end{align}

The integral $J_{c,c}$ contains a function $C(d,\Lambda)$, where $d$ is the
number of space-time dimensions and $\Lambda$ is the mass scale that arises in
dimensional regularization. In the limit $d\rightarrow4$ this function becomes
divergent and needs to be removed by renormalization. We neglect this
contribution because it has zero range and overlaps with other short distance
effects not considered here.

The function $[\Gamma_{N}^{+}]$ is related to the scalar form factor
$\sigma(t)$ by $\langle p^{\prime}|\mathcal{L}_{sb}|p\rangle=-\,\sigma
(t)\,\left[  \bar{u}\,u\right]  $, where $\mathcal{L}_{sb}$ is the symmetry
breaking Lagrangian. Its long range structure, as discussed by Gasser, Sainio
and \v{S}varc \cite{GSS}, is associated with diagrams 2a and 2b, and hence, in
our notation, one has the equivalence
\begin{equation}
\sigma(t)\;[\bar{u}\,u]\;=\;3\,\mu^{2}\;[\Gamma_{N}^{+}]\;, \label{C11}%
\end{equation}
which is valid for large distances. This allows the asymptotic scalar
potential to be written as
\begin{equation}
\mathcal{T}^{S}\;\cong\;2\,\frac{\alpha_{00}^{+}}{\mu}\,\frac{\sigma(t)}%
{\mu^{2}}\;[\bar{u}\,u]^{(1)}\;[\bar{u}\,u]^{(2)}\;. \label{C12}%
\end{equation}

This result is interesting because it sheds light into the structure of the
interaction. The picture that emerges is that of a nucleon, acting as a scalar
source, disturbing the pion cloud of the other. The function $\sigma(t)$ is
related to the $\pi N$ $\sigma$-term by $\sigma(0)=\sigma_{N}$ and its value
at the Cheng-Dashen point $t=2\mu^{2}$ may be extracted from experiment.

In some situations, it may be useful to use an effective parametrized version
of $\sigma(t)$. In this case, the $t$ dependence of Eqs.(\ref{C7}-\ref{C8})
suggests that one should use the form
\begin{equation}
\sigma(t)\;\cong\;-\;\frac{c}{t-m_{s}^{2}}\;, \label{C13}%
\end{equation}
where the free parameters $c$ and $m_{s}$ may be written in terms of
$\sigma(2\mu^{2})$ and $\sigma(0)$ as
\begin{align}
c  &  =\sigma(0)\;m_{s}^{2}\;,\label{C14}\\[0.4cm]
m_{s}^{2}  &  =\frac{2\;\sigma(2\mu^{2})}{\sigma(2\mu^{2})-\sigma(0)}\;\mu
^{2}\,. \label{C15}%
\end{align}

The coupling constant of this effective scalar state to nucleons may be
obtained by comparing Eq.(\ref{C12}) with
\begin{equation}
\mathcal{T}^{S}\;\cong\;-\,\frac{g_{s}^{2}}{t-m_{s}^{2}}\;[\bar{u}%
\,u]^{(1)}\;[\bar{u}\,u]^{(2)} \label{C16}%
\end{equation}
and one has
\begin{equation}
g_{s}^{2}\;\cong\;2\,\alpha_{00}^{+}\;\frac{m_{s}^{2}\,\sigma(0)}{\mu^{3}%
}\;. \label{C17}%
\end{equation}

In Tab.\ref{Tab.1} we display the values of $g_{s}$ and $m_{s}$ obtained from
input factors found in the recent literature. In most cases, the scalar mass
is close to that used in the Bonn potential \cite{B}, but the coupling
constant is smaller. We would like to stress, however, that the purpose of
this exercise is not to predict theses values. Instead, it is to show that the
actual asymptotic exchange of two uncorrelated pions may be naturally
simulated in terms of an effective scalar interaction. As a final comment, one
notes that the coupling constant given by Eq.(\ref{C16}) vanishes in the
chiral limit and hence the effective approach is not equivalent to the linear
$\sigma$-model, in which this does not happen.

\begin{table}[h]
\caption{Predictions for $g_{s}$ and $m_{s}$ from Eqs. (\ref{C14}) and
(\ref{C15}), using the following input parameters: $a\rightarrow
\,$Ref.\cite{H83}, $b\rightarrow\,$Ref.\cite{GSS}, $c\rightarrow
\,$Ref.\cite{GLS}, $d\rightarrow\,$Ref.\cite{Kauf+99} and $e\rightarrow
\,$Ref.\cite{Pav+99}.}%
\label{Tab.1}%
\[%
\begin{tabular}
[c]{ccccc}\hline\hline
$\;\;\;\;\;\;\;\alpha_{00}^{+}\;\;\;\;\;\;\;$ & $\;\;\sigma(2\mu^{2}%
)\;$(MeV)\ \  & $\overset{\;}{\underset{\,}{\left[  \sigma(2\mu^{2}%
)\!-\!\sigma(0)\right]  }}\;$(MeV) & $\;\;\;\;\;\;\;\;g_{s}\;\;\;\;\;\;\;\;$ &
$\;\;\;m_{s}\;$(MeV)$\;\;\;$\\\hline
3.68$\ ^{a}$ & 60\ $^{c}$ & $\overset{\,}{7.3\ ^{b}}$ & 7.22 & 564\\
3.68\ $^{a}$ & 60\ $^{c}$ & 15\ $^{c}$ & 4.36 & 393\\
6.74\ $^{d}$ & 88\ $^{d}$ & 15\ $^{c}$ & 9.09 & 478\\
4.61\ $^{e}$ & 90\ $^{e}$ & $\underset{\,}{15\ ^{c}}$ & 7.71 &
483\\\hline\hline
\end{tabular}
\]
\end{table}

The non relativistic potential in configuration space is
\begin{align}
V^{S}(x)  &  =-\,2\,\alpha_{00}^{+}\;\frac{\mu}{4\pi}\left[  \frac{4\pi}%
{\mu^{4}}\int\frac{d^{3}\Delta}{(2\pi)^{3}}\;e^{-i\,\mathbf{\Delta\,\cdot\,r}%
}\;\sigma(-\mathbf{\Delta}^{2})\right] \nonumber\\[0.5cm]
&  =-\,\left[  3\,\alpha_{00}^{+}\;\frac{\mu}{m}\left(  \frac{g}{4\pi}\right)
^{2}\right]  \,\frac{\mu}{4\pi}\,\left[  S_{c,c}(x)-S_{c,sN}^{(1)}(x)\right]
\;, \label{C18}%
\end{align}
where $x=\mu\,r$ and
\begin{align}
S_{c,c}(x)  &  =\int_{0}^{1}d\alpha\int_{0}^{1}d\beta\;\frac{\lambda^{2}%
}{\beta}\;\frac{e^{-\lambda\,x}}{x}\;,\label{C19}\\[0.6cm]
S_{c,sN}^{(1)}(x)  &  =\frac{2m^{2}}{\mu^{2}}\int_{0}^{1}d\alpha
\;\frac{1-\alpha}{\alpha}\int_{0}^{1}d\beta\;\frac{1-\beta}{\beta}%
\;\frac{e^{-\eta\,x}}{x}\;. \label{C20}%
\end{align}

\begin{figure}
[h]
\begin{center}
\includegraphics[
height=2.9888in,
width=4.0906in
]%
{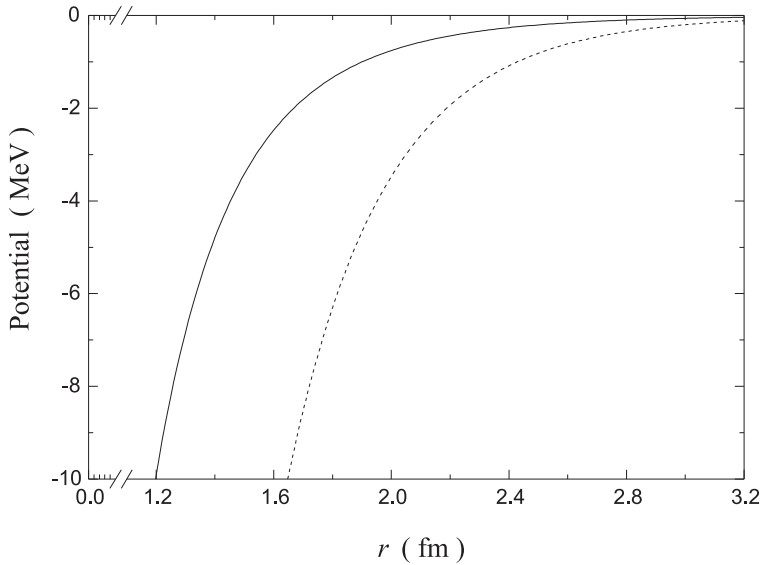}%
\caption{Scalar-isoscalar potential: asymptotic two-pion exchange (solid
line), Eq.(18), and effective scalar exchange (dashed line), Eq.(16), with
$\alpha_{00}^{+}=4.61$, $g_{s}=7.71$ and $m_{s}=483$ MeV taken from the last
line of Tab.1.}%
\label{Fig.3}%
\end{center}
\end{figure}

It is important to note that these functions $S$ are not of the Yukawa type
and hence cannot be represented over their full range by terms proportional to
$e^{-\,m_{s}\,r}/r$, irrespectively of the value chosen for the parameter
$m_{s}$. In Fig.\ref{Fig.3} we display this potential together with that due
to the exchange of an effective scalar meson.%

\section{The Kernel}

\begin{figure}
[h]
\begin{center}
\includegraphics[
height=4.1684in,
width=5.0315in
]%
{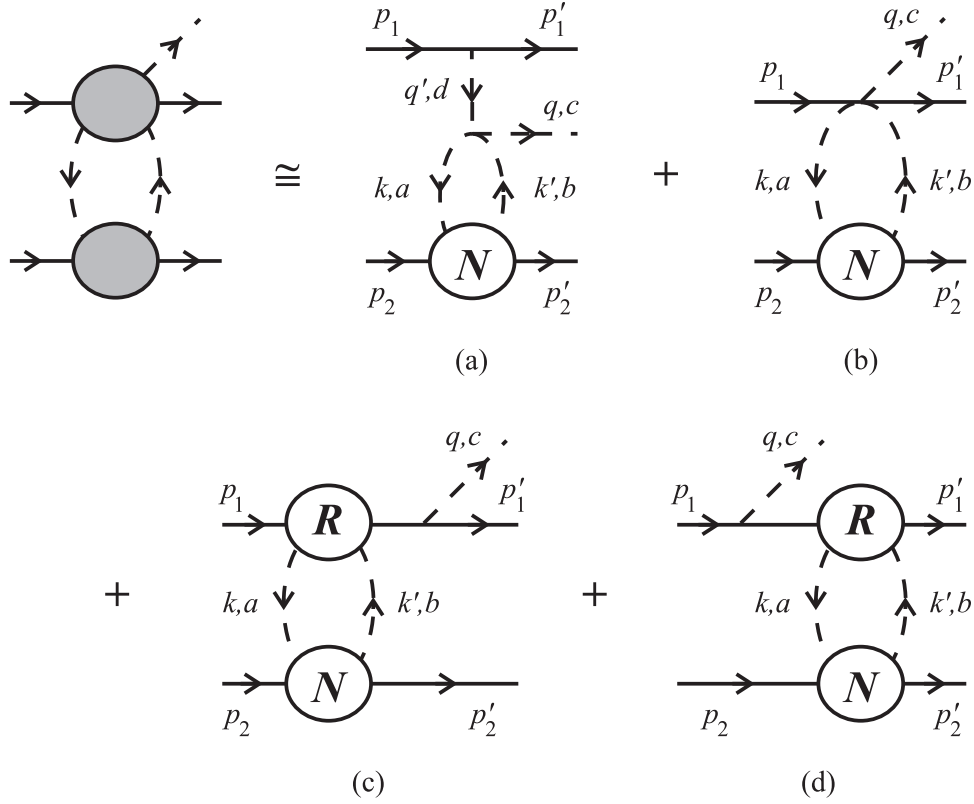}%
\caption{Contributions to the $NN\rightarrow\pi NN$ kernel: (a) pion-pole, (b)
contact, (c) and (d) z-graphs.}%
\label{Fig.4}%
\end{center}
\end{figure}

In this section we construct a kernel for pion production in $NN$ scattering
and due to the exchange of two pions. It is represented in Fig.\ref{Fig.4},
denoted by $\mathcal{T}$ and based on $T_{cba}$ and $T_{ba}$, the amplitudes
for the processes $\pi N\rightarrow\pi\pi N$ and $\pi N\rightarrow\pi N$,
respectively. The kernel $\mathcal{T}$ for an outgoing pion with momentum $q$
and isospin index $c$ is
\begin{equation}
\mathcal{T}_{c}=-\,i\,\frac{1}{2!}\int\frac{d^{4}Q}{(2\pi)^{4}}\frac
{T_{cba}\;T_{ba}}{(k^{2}-\mu^{2})(k^{\prime\,2}-\mu^{2})}\;. \label{T1}%
\end{equation}%

The basic subamplitudes have the isospin structures%
\begin{align}
T_{ba}  &  =\delta_{ab}\,T^{+}+i\,\epsilon_{bac}\tau_{c}\,T^{-}\;,
\label{T2}\\[0.08in]
T_{cba}  &  =-\,i\,\left\{  \delta_{bc}\tau_{a}\,T_{A}+\delta_{ac}\tau
_{b}\,T_{B}+\delta_{ab}\tau_{c}\,T_{C}+i\,\epsilon_{cba}\,T_{E}\right\}
\label{T3}%
\end{align}
and hence
\begin{equation}
\mathcal{T}_{c}=\tau_{c}^{(1)}\,\mathcal{T}_{1}+i\,(\mbox{\boldmath
$\tau$}^{(1)}\times\mbox{\boldmath$\tau$}^{(2)})_{c}\,\mathcal{T}_{12}%
+\tau_{c}^{(2)}\,\mathcal{T}_{2}\;, \label{T4}%
\end{equation}
where
\begin{align}
\mathcal{T}_{1}  &  =-\;\frac{1}{2}\int\frac{d^{4}Q}{(2\pi)^{4}}%
\;\frac{\left[  T_{A}+T_{B}+3\;T_{C}\right]  ^{(1)}\;\left[  T^{+}\right]
^{(2)}}{(k^{2}-\mu^{2})(k^{\prime\,2}-\mu^{2})}\;,\label{T5}\\[2mm]
\mathcal{T}_{12}  &  =-\;\frac{1}{2}\int\frac{d^{4}Q}{(2\pi)^{4}}%
\;\frac{\left[  T_{A}-T_{B}\right]  ^{(1)}\;\left[  T^{-}\right]  ^{(2)}%
}{(k^{2}-\mu^{2})(k^{\prime\,2}-\mu^{2})}\;,\label{T6}\\[2mm]
\mathcal{T}_{2}  &  =-\;\frac{1}{2}\int\frac{d^{4}Q}{(2\pi)^{4}}%
\;\frac{2\;\left[  T_{E}\right]  ^{(1)}\;\left[  T^{-}\right]  ^{(2)}}%
{(k^{2}-\mu^{2})(k^{\prime\,2}-\mu^{2})}\;. \label{T7}%
\end{align}

We begin by discussing the process $\pi^{b}(k^{\prime})\,N(p)\rightarrow
\pi^{a}(k)\,\pi^{c}(q)\,N(p^{\prime})$. The amplitude $T_{cba}$ is given by
the sum of $T_{cba}^{\pi}$, a $t$-channel pion-pole contribution, and a
remainder, denoted by $\bar{T}_{cba}$. The explicit forms of these terms, for
a system containing just pions and nucleons, was presented in Ref.\cite{PR}
and here we just quote the main results.

The pion-pole amplitude for on-shell nucleons is
\begin{equation}
i\,T_{cba}^{\pi}=-\;\frac{m\,g_{\mathrm{A}}}{f_{\pi}}\;\left[  \bar{u}%
\,\tau_{d}\,\gamma_{5}\,u\right]  \;\frac{T_{dcba}^{\pi\pi}}{(p^{\prime
}\!-\!p)^{2}-\mu^{2}}\;, \label{T8}%
\end{equation}
where $f_{\pi}$ and $g_{\mathrm{A}}$ are the pion and axial decay constants,
whereas $T_{dcba}^{\pi\pi}$ is the pion scattering amplitude. At tree level,
it is given by
\begin{equation}
T_{dcba}^{\pi\pi}=\frac{1}{f_{\pi}^{2}}\;\left\{  \delta_{ad}\delta
_{bc}\left[  (q\!-\!k^{\prime})^{2}\!-\!\mu^{2}\right]  \!+\!\delta_{bd}%
\delta_{ac}\left[  (q\!+\!k)^{2}\!-\!\mu^{2}\right]  \!+\!\delta_{cd}%
\delta_{ab}\left[  (k^{\prime}\!-\!k)^{2}\!-\!\mu^{2}\right]  \right\}
\label{T9}%
\end{equation}
and one has
\begin{equation}
T_{A}^{\pi}=-\;\frac{m\,g_{\mathrm{A}}}{f_{\pi}^{3}}\;\frac{(p^{\prime
}\!-\!p\!+\!k)^{2}-\mu^{2}}{(p^{\prime}\!-\!p)^{2}-\mu^{2}}\;. \label{T10}%
\end{equation}

The evaluation of $\bar{T}_{cba}$ requires the calculation of a large number
of diagrams. However, long ago Olsson and Turner \cite{OT} have shown that its
leading contribution comes from the effective Lagrangian
\begin{equation}
\bar{\mathcal{L}}=\frac{g_{\mathrm{A}}}{8\,f_{\pi}^{3}}\;\bar{\psi}%
\,\gamma_{\mu}\,\gamma_{5}\,\mbox{\boldmath$\tau$}\,\psi\cdot\mbox{\boldmath
$\phi$}\,\partial^{\mu}\mbox{\boldmath$\phi$}^{2}\;, \label{T11}%
\end{equation}
which gives the following contribution to $\bar{T}_{A}$
\begin{equation}
\bar{T}_{A}=\frac{2\,g_{\mathrm{A}}}{8\,f_{\pi}^{3}}\left(  2m+%
\!\!\not\!%
k\right)  \;. \label{T12}%
\end{equation}

The corresponding expressions for $T_{B}$ and $T_{C}$ are obtained by making
$k\rightarrow-k^{\prime}$ and $k\rightarrow q$, respectively.

The main implication of this structure of the $\pi N\rightarrow\pi\pi N$
interaction for our study is that the leading contribution to $\mathcal{T}$
comes from the diagrams 4a and 4b. As the $NN$ interaction due to the exchange
of two pions is dominated by the scalar-isoscalar channel, in this work we
consider only the amplitude $\mathcal{T}_{1}$, Eq.(\ref{T5}), and postpone the
discussion of the remaining components to another occasion.

Diagram 4a yields
\begin{align}
\mathcal{T}_{1}^{\pi}  &  =\left[  \frac{m\,g_{\mathrm{A}}}{f_{\pi}^{3}%
}\right]  \frac{1}{(p_{1}^{\prime}\!-\!p_{1})^{2}-\mu^{2}}\;[\bar{u}%
\,\gamma_{5}\,u]^{(1)}\nonumber\\[0.3cm]
&  \times\frac{1}{2}\int\frac{d^{4}Q}{(2\pi)^{4}}\;\frac{\left[
2(p_{1}^{\prime}\!-\!p_{1})^{2}+2q\!\cdot\!\Delta+3\Delta^{2}/2-5\mu
^{2}+2Q^{2}\right]  \;\left[  T^{+}\right]  ^{(2)}}{[(Q\!-\!\Delta
/2)^{2}\!-\!\mu^{2}]\;[(Q\!+\!\Delta/2)^{2}\!-\!\mu^{2}]}\;. \label{T13}%
\end{align}

We note that the last two terms in the result $Q^{2}=(\mu^{2}-\Delta
^{2}/4)+[(Q\!-\!\Delta/2)^{2}\!-\!\mu^{2}]/2+[(Q\!+\!\Delta/2)^{2}\!-\!\mu
^{2}]/2$ allow the cancellation of pion propagators and therefore correspond
to short range effects that will be neglected here. We then obtain
\begin{equation}
\mathcal{T}_{1}^{\pi}=i\,\left[  \frac{m\,g_{\mathrm{A}}}{f_{\pi}^{3}}\right]
\,\left\{  3+\frac{3q^{2}+4q\!\cdot\!(p_{1}^{\prime}\!-\!p_{1})}%
{(p_{1}^{\prime}\!-\!p_{1})^{2}-\mu^{2}}\right\}  \,[\bar{u}\,\gamma
_{5}\,u]^{(1)}\;\left[  \Gamma_{N}^{+}\right]  ^{(2)}\,, \label{T14}%
\end{equation}
where $\left[  \Gamma_{N}^{+}\right]  ^{(2)}$ is given by Eq.(\ref{C5}). This
result may be associated with the scattering of a pion emitted in one of the
nucleons by the pion cloud of the other, indicating that the kernel one is
considering here is not fully disentangled from that usually called pion
rescattering. Indeed, the description of the rescattering process is based on
an intermediate $\pi N$ amplitude for off-shell pions, which satisfies a
Ward-Takahashi identity \cite{WT}. In the isospin symmetric channel, this
identity may be expressed as
\begin{equation}
T^{+}(q^{\prime\,2},q^{2})=T_{N}^{+}\;+\;\frac{q^{\prime\,2}+q^{2}-\mu^{2}%
}{f_{\pi}^{2}\,\mu^{2}}\;\sigma(t)\;[\bar{u}\,u]\;+\;r^{+}\;, \label{T15}%
\end{equation}
where $T_{N}^{+}$ is the nucleon pole (Born) term evaluated with pseudovector
coupling, $q$ and $q^{\prime}$ are the momenta of the pions, $\sigma(t)$ is
the scalar form factor and $r^{+}$ is a remainder that does not include
leading order contributions.

The only term that depends strongly on off-shell effects is that proportional
to the scalar form factor and hence one writes
\begin{equation}
T^{+}(q^{\prime\,2},q^{2})=T^{+}(\mu^{2},\mu^{2})+\delta T^{+}\,, \label{T16}%
\end{equation}
with
\begin{equation}
\delta T^{+}=\frac{(q^{\prime\,2}-\mu^{2})+(q^{2}-\mu^{2})}{f_{\pi}^{2}%
\,\mu^{2}}\;\sigma(t)\;[\bar{u}\,u]\;. \label{T17}%
\end{equation}

The contribution of this factor to the pion rescattering amplitude on nucleon
2 reads
\begin{equation}
\mathcal{T}_{1}^{\delta}=i\,\left[  \frac{m\,g_{\mathrm{A}}}{f_{\pi}^{3}%
}\right]  \,\left\{  3+\frac{3(q^{2}-\mu^{2})}{(p_{1}^{\prime}-p_{1})^{2}%
-\mu^{2}}\right\}  \,\left[  \bar{u}\,\gamma_{5}\,u\right]  ^{(1)}\;\left[
\Gamma_{N}^{+}\right]  ^{(2)}\,, \label{T18}%
\end{equation}
using Eq.(\ref{C11}). Adding this result to the on-shell $\pi N$ amplitude
derived by Gasser, Sainio and \v{S}varc, Ref.\cite{GSS}-Eq.(A.35), one
recovers Eq.(\ref{T14}). Therefore, in the sequence, we no longer consider the
term proportional to the pion-pole in that expression, with the understanding
that it should be included in the on-shell rescattering amplitude.

The evaluation of diagram 4b is more straightforward and produces
\begin{equation}
\bar{\mathcal{T}}_{1}=-\,i\,\left[  \frac{m\,g_{\mathrm{A}}}{f_{\pi}^{3}%
}\right]  \,\left[  \bar{u}\,\left\{  2+\frac{%
\!\!\not\!%
q}{2m}\right\}  \,\gamma_{5}\,u\right]  ^{(1)}\,\left[  \Gamma_{N}^{+}\right]
^{(2)}\,. \label{T19}%
\end{equation}

Using the Goldberger-Treiman relation and Eq.(\ref{C11}), we have
\begin{equation}
\mathcal{T}_{1}^{\pi}+\bar{\mathcal{T}}_{1}=i\,\left[  \frac{g}{3\mu^{2}%
f_{\pi}^{2}}\right]  \sigma(t)\left[  \bar{u}\,\left\{  1-\;\frac{%
\!\!\not\!%
q}{2m}\right\}  \,\gamma_{5}\,u\right]  ^{(1)}\,\left[  \bar{u}\,u\right]
^{(2)}\,. \label{T20}%
\end{equation}

One notes that when $\mathcal{T}_{1}^{\pi}$ and $\bar{\mathcal{T}}_{1}$ are
added together, a cancellation occurs, which springs from the same mechanism
and is very similar to that noticed long ago in the study of exchange currents
in pion-deuteron scattering \cite{RW}.

Another contribution with the same two-pion range comes from diagrams 4c and
4d, which yield
\begin{equation}
\mathcal{T}_{1}^{z}=i\,\left[  \frac{g\,\alpha_{00}^{+}}{m\,\mu^{3}}\right]
\,\sigma(t)\,\left[  \bar{u}\,\left\{  1-\;\frac{m\;%
\!\!\not\!%
q}{(p^{\prime}\!+\!q)^{2}-m^{2}}-\;\frac{m\;%
\!\!\not\!%
q}{(p\!-\!q)^{2}-m^{2}}\right\}  \,\gamma_{5}\,u\right]  ^{(1)}\,\left[
\bar{u}\,u\right]  ^{(2)}\,. \label{T21}%
\end{equation}

This result includes the propagation of positive energy states, that do not
contribute to the kernel. Eliminating them and neglecting small non covariant
terms, we have
\begin{equation}
\mathcal{T}_{1}^{z}=i\,\left[  \frac{g\,\alpha_{00}^{+}}{m\,\mu^{3}}\right]
\,\sigma(t)\,\left[  \bar{u}\,\left\{  1+\frac{%
\!\!\not\!%
q}{2m}\right\}  \,\gamma_{5}\,u\right]  ^{(1)}\,\left[  \bar{u}\,u\right]
^{(2)}\,. \label{T22}%
\end{equation}

Our final expression for the covariant kernel is obtained by adding
Eqs.(\ref{T20}) and (\ref{T22}) and reads
\begin{equation}
\mathcal{T}_{1}=i\,\sigma(t)\,g\,\left\{  \left[  \frac{1}{3\,\mu^{2}\,f_{\pi
}^{2}}+\frac{\alpha_{00}^{+}}{m\,\mu^{3}}\right]  \,\bar{u}\,\gamma
_{5}\,u-\left[  \frac{1}{3\,\mu^{2}\,f_{\pi}^{2}}-\frac{\alpha_{00}^{+}%
}{m\,\mu^{3}}\right]  \,\bar{u}\,\frac{%
\!\!\not\!%
q}{2m}\,\gamma_{5}\,u\right\}  ^{(1)}\,\left[  \bar{u}\,u\right]  ^{(2)}\,.
\label{T23}%
\end{equation}

This covariant amplitude is our main result.


\section{Application}

In order to consider applications in low energy processes, we perform a
non-relativistic approximations in our results. In the case of the central
potential, Eq.(\ref{C11}), one has
\begin{equation}
t^{S}\,=\,2\,\alpha_{00}^{+}\,\sigma(t)\,/\,\mu^{3}\;, \label{A1}%
\end{equation}
where $t=-\,\mathbf{\Delta}^{2}$ and we have discarded a normalization factor
$4m^{2}$. On the other hand, the kernel, suited to be used with nuclear wave
functions, is
\begin{equation}
t_{1}=i\;\sigma(t)\;\frac{g}{2m}\left\{  -\left[  \frac{1}{3\,\mu^{2}\,f_{\pi
}^{2}}+\frac{\alpha_{00}^{+}}{m\,\mu^{3}}\right]  \mbox{\boldmath$\sigma
$}\!\cdot\!(\mbox{\boldmath$p$}^{\prime}-\mbox{\boldmath$p$})+\left[  \frac
{1}{3\,\mu^{2}\,f_{\pi}^{2}}-\frac{\alpha_{00}^{+}}{m\,\mu^{3}}\right]
\mbox{\boldmath$\sigma$}\!\cdot\!\mbox{\boldmath$q$}\right\}  ^{(1)}.
\label{A2}%
\end{equation}

The term proportional to $\mbox{\boldmath$\sigma$}\!\cdot\!(\mbox
{\boldmath$p$}^{\prime}-\mbox{\boldmath$p$})/3\mu^{2}f_{\pi}^{2}$ in this
result coincides with that produced recently in Ref.\cite{BKM}.

\begin{figure}
[h]
\begin{center}
\includegraphics[
height=3.045in,
width=4.414in
]%
{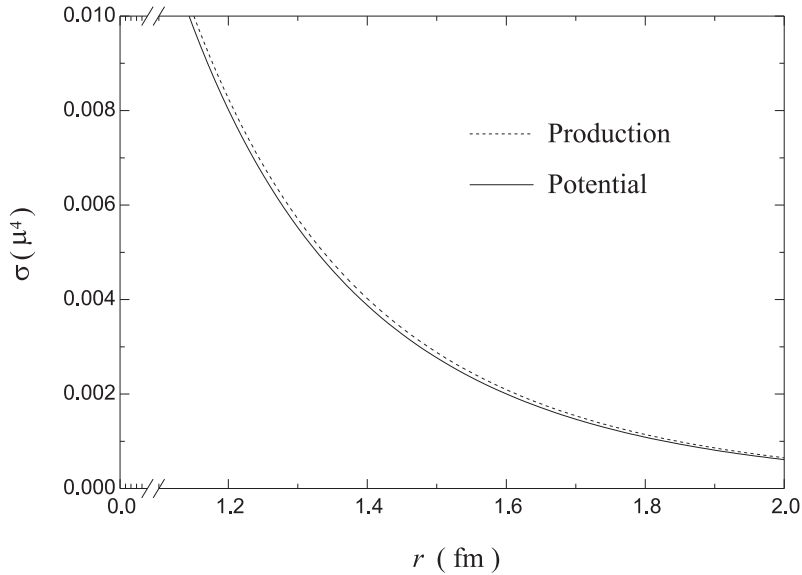}%
\caption{Fourier transform of $\sigma(t)$, that determine the space dependence
of the three body force and production kernels, as function of the distance
$r$.}%
\label{Fig.5}%
\end{center}
\end{figure}

As discussed in Refs.\cite{CPR} and \cite{MRtbf}, a pion production kernel
such as $t_{1}$ gives rise to three body forces and again one has
$t\cong-\,\mathbf{\Delta}^{2}$. In the case of threshold pion production, on
the other hand, $t\cong\mu^{2}/4-\mathbf{\Delta}^{2}$. In order to test the
influence of these different values of $t$, in Fig.\ref{Fig.5} we plot the
Fourier transform of the function $\sigma(t)$, that dictates the space
dependence of the kernel in the two cases. Inspecting it, one learns that the
energy component of the four momentum transferred has little importance and
hence the static result also holds for the production kernel. This allows one
to relate it directly to the central potential
\begin{equation}
t_{1}=i\,\frac{g_{\mathrm{A}}}{f_{\pi}}\,\frac{t^{S}}{2m}\left\{  -\,\left[
\frac{\mu\,m}{6\,\alpha_{00}^{+}\,f_{\pi}^{2}}+\frac{1}{2}\right]
\,\mbox{\boldmath
$\sigma$}\!\cdot\!(\mbox{\boldmath$p$}^{\prime}-\mbox{\boldmath$p$})+\left[
\frac{\mu\,m}{6\,\alpha_{00}^{+}\,f_{\pi}^{2}}-\frac{1}{2}\right]
\,\mbox{\boldmath$\sigma$}\!\cdot\!\mbox{\boldmath$q$}\right\}  ^{(1)}.
\label{A3}%
\end{equation}%

In order to use these results in actual calculations, in either momentum or
configuration spaces, one has to evaluate the function $\sigma(t)$ numerically
and then, the sandwich of the kernel between two-nucleon wave functions. Since
the kernel and the central potential are closely related, consistency would
require that the same dynamics should be used in the construction of both the
operator $t_{1}$ and the wave functions. However, at present, the potential
due to the exchange of two pions is reliable at large distances only and hence
it is not suited to determine wave functions by means of the Schr\"{o}dinger
equation. Therefore, the possibility of using Eq.(\ref{A3}) with ones
favourite scalar potential is an interesting one.


\section{Summary and Conclusions}

In this work we have shown that the central component of the $NN$ potential at
large distances, which is due to the exchange of two uncorrelated pions, may
be naturally expressed in terms of $\sigma(t)$, the scalar form factor. This
function is related to the $\pi N$ $\sigma$-term and may, if one wishes, be
parametrized as an effective scalar meson exchange. However, the coupling of
this state to nucleons vanishes in the chiral limit and hence this scalar
meson does not correspond to that present in the linear $\sigma$-model.

We have also obtained a two-pion-exchange kernel for the process
$NN\rightarrow\pi NN$, that can be applied in both three body forces and pion
production in $NN$ scattering. The complete calculation of this kernel would
require the evaluation of a large number of diagrams. Thus, in order to
estimate the dominant contribution at large $NN$ distances, we have used just
the leading contributions to the subamplitudes $\pi N\rightarrow\pi N$ and
$\pi N\rightarrow\pi\pi N$, in the framework of chiral symmetry. The
simplified result so obtained involves a cancellation between
contact-three-pion and pion-pole vertices. The latter may also be associated
with an off-shell intermediate $\pi N$ amplitude and has been include into the
Tucson-Melbourne \cite{TM} two-pion exchange three nucleon potential. This
means that this force does include a term describing a two-pion exchange
between a pair of nucleons. Thus, the use of an on-shell $\pi N$ amplitude
gives rise to a less ambiguous definition of the two-pion exchange three body
force \cite{BR}.

At large distances, the kernel is closely related to the two-pion exchange
scalar isoscalar $NN$ potential. Indeed, in the case of three body forces, we
could show that the kernel and the potential have the same spatial dependence.
For threshold pion-production, this relationship is also approximately valid.
These results led us to produce expressions that relate directly the kernel to
the potential. Using the extreme numerical values for $\alpha_{00}^{+}$ found
in the literature, namely 3.68 \cite{H83} and 6.74 \cite{Kauf+99}, in
Eq.(\ref{A3}), one has%
\begin{equation}
t_{1}=\,i\;\frac{g_{\mathrm{A}}}{f_{\pi}}\,\frac{t^{S}}{2m}\left\{
-\,1.18\,\mbox{\boldmath$\sigma$}\!\cdot\!(\mbox{\boldmath$p$}^{\prime}%
-\mbox{\boldmath$p$})+0.18\,\mbox{\boldmath$\sigma$}\!\cdot\!\mbox
{\boldmath$q$}\right\}  ^{(1)}\label{R1}%
\end{equation}
or%
\begin{equation}
t_{1}=\,i\;\frac{g_{\mathrm{A}}}{f_{\pi}}\,\frac{t^{S}}{2m}\left\{
-\,0.87\,\mbox{\boldmath$\sigma$}\!\cdot\!(\mbox{\boldmath$p$}^{\prime}%
-\mbox{\boldmath$p$})-0.13\,\mbox{\boldmath$\sigma$}\!\cdot\!\mbox
{\boldmath$q$}\right\}  ^{(1)}\,.\label{R2}%
\end{equation}

These results are quite close to the kernel obtained by ourselves sometime ago
\cite{MRtbf}, given by
\begin{equation}
t_{1}=\,i\;\frac{g_{\mathrm{A}}}{f_{\pi}}\,\frac{t^{S}}{2m}\,\left\{
-\,\mbox{\boldmath
$\sigma$}\!\cdot\!(\mbox{\boldmath$p$}^{\prime}-\mbox{\boldmath$p$})\right\}
^{(1)}\,, \label{R3}%
\end{equation}
in the case of a models based on effective scalar-isoscalar
mesons.\footnote{In that work we have used $g_{A}=1$.} This allows one to
consider the relationship between the kernel and the potential to be a rather
general one. The reason for this generality springs from the old insight by
Nambu \cite{Nambu} and Weinberg \cite{Wpi} that, for generic states $A$ and
$B$, the leading contributions to the process $A\rightarrow\pi B$ are obtained
by inserting the pion, with gradient coupling, into the external lines of the
process $A\rightarrow B$.

Finally, we would like to point out that we may expect the contributions from
the kernel $t_{1}$ to be large. In order to see this, note that momentum
conservation allows one to write
\begin{equation}
t_{1}=\,i\;\frac{g_{\mathrm{A}}}{f_{\pi}}\,\frac{t^{S}}{2m}\,\left\{
\mbox{\boldmath
$\sigma$}\!\cdot\!(\mbox{\boldmath$q$}-\mbox{\boldmath$\Delta$})\right\}
^{(1)} \label{R4}%
\end{equation}
and, in the case of threshold pion production, in configuration space one has
\begin{equation}
t_{1}=\,\frac{g_{\mathrm{A}}}{f_{\pi}}\,\frac{\mu}{2m}\;\mbox{\boldmath
$\sigma$}^{(1)}\!\cdot\!\mbox{\boldmath$\nabla$}_{x}\,V^{S}(x)\;. \label{R5}%
\end{equation}

As the central potential contains Yukawa functions with effective masses which
are not small, its gradient produces a large kernel, proportional to those masses.\newpage%

\baselineskip=0.5cm

\noindent\textbf{Acknowledgements}

We thank Bira van Kolck and Carlos A. da Rocha for useful conversations.
C.M.M. thanks the hospitality of the Instituto de F\'{\i}sica Te\'{o}rica
(Universidade Estadual Paulista) in the initial stage of this work and
acknowledges the support of FAPESP (Brazilian Agency, grant 99/00080-5) and
NSF (grant PHY\_94-20470). J.C.P acknowledges the support of FAPESP (Brazilian
Agency, grant 94/03469-7).%

\end{document}